# Controlling ion kinetic energy distributions in laser produced plasma sources by means of a picosecond pulse pair

Aneta S. Stodolna[1], Tiago de Faria Pinto[1], Faisal Ali[1], Alex Bayerle[1], Dmitry Kurilovich[1,2], Jan Mathijssen[1], Ronnie Hoekstra[1,3], Oscar O. Versolato[1], Kjeld S. E. Eikema[1,2], Stefan Witte[1,2,*]

[1] Advanced Research Center for Nanolithography (ARCNL), Science Park 110, 1098 XG Amsterdam, The Netherlands
[2] Department of Physics and Astronomy, Vrije Universiteit, De Boelelaan 1081, 1081 HV Amsterdam, The Netherlands
[3] Zernike Institute for Advanced Materials, University of Groningen, Nijenborgh 4, 9747 AG Groningen, The Netherlands
[*] Email: witte@arcnl.nl

**The next generation of lithography machines uses extreme ultraviolet (EUV) light originating from laser-produced plasma (LPP) sources, where a small tin droplet is ionized by an intense laser pulse to emit the requested light at 13.5 nm. Numerous irradiation schemes have been explored to increase conversion efficiency (CE), out of which a double-pulse approach comprising a weak picosecond Nd:YAG pre-pulse followed by a powerful pulse is considered to be very promising [1]. Nevertheless, even for such CE-optimized schemes, ion debris ejected from the plasma with kinetic energies up to several keV remain a factor that hampers the maximum performance of LPP sources. In this letter we propose a novel pre-pulse scheme consisting of a picosecond pulse pair at 1064 nm, which decreases the amount of undesirable fast ions, avoids back-reflections to the lasers and enables one to tailor the target shape.**

In the past two decades a large number of theoretical and experimental studies have been conducted on possible light sources for EUV lithography, including synchrotron radiation [2, 3], free-electron lasers [4, 5], plasma sources [6-10] and high-harmonic generation [11]. From the aforementioned solutions, a tin-based laser-produced plasma source received the most attention due to its high conversion efficiency, robustness and scalability [12, 13], resulting in a first commercial machine launched in 2010. In such an LPP source, narrowband radiation around 13.5 nm comes from multiple ionic states, $Sn^{8+}$ to $Sn^{14+}$ [14, 15], collisionally excited by plasma electrons heated through interaction with a powerful $CO_2$ laser. An effective coupling between laser light and plasma occurs near the critical density, which for $CO_2$-laser-driven plasma is around $10^{19}$ cm$^{-3}$, meaning that a mass-limited tin target should be expanded to reduce its density. At the same time the size of the EUV source cannot be too large to match the requirements for the maximum etendue [16]. The precise control of the target shape is thus crucial for the production of EUV light in an industrial setting.

The expansion of the target can be achieved by deforming the tin droplet with a pre-pulse generated either by the same $CO_2$ laser system [17, 18] or by a separate laser, typically Nd:YAG [19-24]. The latter solution reduces the amount of backscattered light and by decoupling both laser systems it prevents instabilities and potential damage to the lasers, at the expense of added complexity in the EUV lithography machine. The interaction between a tin droplet and a nanosecond pre-pulse leads to the generation of a high-density disk target and results in a reported conversion efficiency of 4.7% [25]. Alternatively, a picosecond pre-pulse could be employed that expands the droplet to a low-density diffuse target resembling an acorn [26, 27], which has a significantly higher surface to volume ratio compared to a disk target, ensuring improved light absorption and, consequently, higher CE with the maximum reported value of 6% [1].

Due to the interaction with intense laser pulses, the source emits large amounts of energetic particles. Out of this debris, the ions with kinetic energies of several keV are particularly undesirable, as they may damage the nearby multi-layer mirror that collects the light emitted by the plasma, reducing its reflectivity and thus limiting its lifetime [28]. This issue is particularly relevant when using picosecond-duration pre-pulses, which are associated with an increase in the emission of ions with multi-keV energy [29]. To mitigate the impact of ion debris several techniques have been introduced including stopping fast ions using a buffer gas [30, 31], or to guide them away to a "dump" using a magnetic field [32], or a combination of both [33].

Alternatively, it may be possible to control the physical mechanism responsible for the acceleration of the produced ions to the observed high velocities. Some prior studies hint towards the feasibility of such an approach. For example, in experiments on solid tin and gadolinium targets [34, 35], the ion energy distributions were shifted significantly towards lower values. This substantial reduction of ions kinetic energy was achieved by using a pulse pair comprising a weak picosecond pulse at different wavelengths (1064 nm, 532 nm or 355 nm) followed by a strong nanosecond pulse at 1064 nm. Recently, a similar observation has been made on droplets in a double pulse irradiation scheme comprising a 7.5 ns Nd:YAG pulse with energy of 48 mJ followed by a 600 mJ $CO_2$ pulse [36]. A maximum reduction in the ion average kinetic energy by a factor of 3 was observed by delaying pulses by 164 ns.

The aforementioned experiments addressed only the influence of a plasma generated by the first pulse (pre-pulse) on the ion energy distribution originated from the second pulse (main pulse). However, in the industrially relevant case a pre-pulse is employed to fluid-dynamically transform the droplet into an optimal target shape. Thus, it is an open question if multi-pulse fast-ion-mitigation schemes are applicable in the production of the acorn-like mass-limited target shapes that are required for high-CE LPP sources. In this letter we address the use of a carefully designed picosecond pulse pair as a pre-pulse to reduce the amount of fast ions and additionally to transform the droplet into the preferred acorn-shaped target.

The experimental setup, shown schematically in Fig. 1, comprised a tin droplet generator operated at approximately 10 kHz repetition rate resulting in 30 µm diameter droplets (a detailed description of the droplet generator is given in Ref. [21]). The droplets were irradiated with a picosecond pulse pair, generated by a home-built Nd:YAG laser similar to the systems presented in Ref. [37-39]. It consisted of a vanadate (Nd:YVO$_4$) oscillator generating 1064 nm pulse trains at a 100 MHz repetition rate, with a pulse duration tunable in the range from 15 ps to 100 ps. In the experiments the pulse duration was measured by means of autocorrelation [40]. A fiber-coupled pulse picking system comprising an acousto-optic modulator in combination with an electro-optic modulator was employed to select two pulses of the same duration at a chosen time delay Δτ, ranging from 10 ns to 1000 ns in increments of 10 ns. This pulse-picking system also reduced the repetition rate to 10 Hz to match the data acquisition rate during the experiments. The selected pulse pair was first pre-amplified by approximately seven orders of magnitude in a bounce amplifier making use of two high-gain Nd:YVO$_4$ crystals, which were side-pumped with diode lasers at 880 nm. Finally, the pulse pair was sent through a post-amplifier containing two Nd:YAG rods resulting in a maximum single pulse energy of 200 mJ. In the experiments the energy of the pulse pair was controlled by a combination of a half-wave plate λ/2 and a thin film polarizer (TFP). The pulse energy of the second, stronger pulse was kept constant at 5 mJ whereas the energy in the first pulse was varied from 0 to 500 µJ with a Pockels cell. Prior to entering a vacuum chamber, a quarter-wave plate provided a circular polarization and a 60 cm lens focused pulses to 135 µm (1/e$^2$) at the position of the droplet. To detect the ions kinetic energy distributions two commercial Faraday cups (Kimball Physics, model FC-73A) were mounted at 30° and 62° with respect to the incident laser beam [41, 42]. The evolution of tin droplets after the interaction with the pulse pair was recorded by means of shadowgraph images obtained from a CCD cameras positioned in the horizontal plane at 90° and 150° along the laser propagation axis, allowing for a side and back view, respectively, using pulsed backlighting at 560 nm wavelength.

A series of shadowgraphs in Fig. 2 demonstrates the influence of picosecond pulses on the droplet deformation and expansion at 550 ns after the laser impact. Fig. 2(a) represents a typical acorn-like shape of a droplet deformed with a single 52 ps pulse (5 mJ) which is composed of two unequal conjunct spheroids resulting from a shock wave propagation [23, 24]. In brief, when an ultra-short laser pulse irradiates a tin droplet within a short amount of time (<1 ns) the light gets absorbed in a thin layer near the surface. This rapid energy deposition gives rise to a hemispherical shock wave, which focuses inside the droplet leading to cavitation and creation of the shell on the front (right) side. The second shell on the rear (left) side results from the spallation effect caused by the shock wave reflected at the back surface.

In contrast, when the same 5 mJ picosecond pulse is preceded by a weak pulse the droplet shape changes noticeably. Remarkably, a 25 µJ pulse proceeding the second pulse by 10 ns flattens the target at the front side due to plasma "push" and reduces the target expansion by 20% along the laser axis (Fig. 2(h)). By doubling the energy in the first pulse to 50 µJ and keeping the time delay at 10 ns, more plasma is being generated, which surrounds the droplet and limits its expansion at the backside as well as in the vertical direction (Fig. 2(b)). For this compressed target a significant reduction in spallation is observed. This may be particularly beneficial for use in LPP sources as such spalling is detrimental for machine lifetime. For longer time delays the plasma gets less intense, particularly at the back side of the target, which spreads more in the vertical direction (Fig. 2(c) and (d)). For time delays longer than 200 ns the target shape reverts back to the acorn-like shape, except at the front side which stays flat as a result of a plasma generated by parasitic pulses present in the pulse train due to limited contrast of the laser setup for time delays $\Delta\tau$ > 100 ns.

When the energy in the first pulse is further increased, no additional compression of the target is observed. In contrast, shadowgraphs taken for 10 ns delay show expansion in the direction of the laser light (Fig. 2(e) and (i)). A possible explanation to this observation might be a shift of the position of critical density away from the droplet in the direction of the laser light with the increase of energy in the 1$^{st}$ pulse. Consequently, the laser light from the 2$^{nd}$ pulse generates plasma further away from the target permitting larger expansion of the droplet in the laser direction. A similar explanation may be used to describe the flattening of the target's front, which is visible in Fig. 2(g) and (j). By increasing the time delay beyond 10 ns, plasma generated by the 1$^{st}$ pulse has more time to fade away and the position of critical density shifts back to the vicinity of the droplet. Thus, the 2$^{nd}$ pulse produces plasma closer to the droplet, which experiences a stronger push at the front.

Different combinations of the first pulse energy ($E_1$) and time delay between two pulses ($\Delta\tau$) result in diverse target shapes. Nevertheless, shadowgraphs in Fig. 2 marked with a thick frame (i.e. images (d), (f) and (h)) reveal the close resemblance of some of these target shapes to the original acorn-like shape. However, as will be discussed below, the ions produced by these pulse pairs have significantly lower kinetic energies.

Figure 3(a) shows the total charge emitted in the direction of the 30° FC for, which was calculated according to:

$$\frac{dQ}{d\varepsilon} = \frac{t^3}{mL^2} \cdot \frac{I(t)}{\Omega X} \quad , \tag{1}$$

where $t$ is the time-of-flight, $I(t)$ the ion current obtained by correcting the measured voltage signals for the response function of the read-out network [41], $m$ the mass of tin, $L$ the time-of-flight distance, $\Omega$ a solid angle and $X$ is a FC grid transmission. As a reference, experiments with a single pulse were performed, where the pulse energy was set to 5 mJ and the pulse duration to 52 ps (black symbols). The colored symbols correspond to measurements with pulse pairs for varying energy in the first pulse and a fixed time delay of 10 ns. The inset shows the number of collected ions grouped in four energy ranges. Remarkably, in the case of a pulse pair in which a first pulse with only 25 µJ energy precedes the stronger 5 mJ pulse (cyan symbols), the ion energy spectrum already shifts towards lower energies and the number of fast ions in the range 3-10 keV decreases by roughly 35% compared to the single pulse case. For $E_1$ = 50 µJ (green symbols) the spectrum changes even more and the measurements demonstrate a one order of magnitude reduction in the amount of the ions with kinetic energy above 3 keV. However, the effect seems to saturate for energies in the 1$^{st}$ pulse above 150 µJ (red symbols) leading to the maximum reduction of fast ions with kinetic energies above 1 keV. In contrast, the number of ions at low energies from 100-300 eV increases by one order of magnitude, whereas in the energy range 0.3-1 keV this increase is only by a factor of 1.3. This growth can be explained by a geometric effect due a mismatch between the laser beam diameter (135 µm) and a tin droplet diameter (30 µm). The first pulse interacts with a droplet and generates plasma which expands well beyond 30 µm within 10 ns, resulting in a bigger target interaction area for the 2$^{nd}$ pulse. Our experiments on a solid tin target (manuscript in preparation), where the beam size matched the target size, show only 14% increase in the total amount of ions produced by a pulse pair compared to a single pulse. Therefore, by matching the beam size with the droplet, a strong reduction in the amount of slow ions as well as further decrease in the number of fast ions can be expected. Furthermore, experiments on the solid target show that the absolute amount of energy in the 1$^{st}$ pulse determines the deceleration effect, not the percentage relation with respect to the 2$^{nd}$ pulse. Therefore, to induce a significant shift to the ion kinetic energy distribution, the 1$^{st}$ pulse needs to create a sufficiently dense plasma. Figure 3(b) shows that by keeping constant energies in both pulses, here E$_1$=150 µJ and E$_2$ =5 mJ, and by delaying the pulses beyond 10 ns (colored symbols) further reduction in the ion kinetic energy is not achieved. Instead it results in an increase in the number of slow ions.

In the single pulse case, the ion energy distribution can be explained by a self-similar model of free plasma expansion into a vacuum based on a hydrodynamic approach [43]. The applicability of this model for ion spectra resulting from an adiabatically expanding plasma has been recently experimentally confirmed by Bayerle *et.al* [41]. The model assumes that initially a plasma occupies the half-space x<0, and that the ions are cold and at rest with a step density function whereas the electrons obey a Boltzmann distribution. Once the plasma starts to expand the ions get accelerated in the electrostatic potential and the number of ions per unit energy and unit surface is given by [44]:

$$\frac{dN}{dE} = \left(n_{i0} c_s t / \sqrt{2\mathcal{E}\mathcal{E}_0}\right) exp\left(-\sqrt{2\mathcal{E}/\mathcal{E}_0}\right) \quad , \tag{2}$$

with $n_{i0}$ being the initial ion density, $c_s$ - the ion-acoustic velocity and $\mathcal{E}_0$ – the characteristic ion energy related to the electrons temperature $T_e$ via $\mathcal{E}_0 = Z k_B T_e$ , where Z is the ion charge number and $k_B$ is the Boltzmann constant.

The dashed black line in Fig. 3(b) shows the fit of the ion kinetic energy spectrum to Eq. (2) for the single pulse case. According to the model, the plasma produced by a 52 ps single pulse at 5 mJ leads to the generation of ions with the characteristic energy of ε$_0$=990 (50) eV. The energy spectra recorded for pulse pairs (colored symbols) show a non-monotonic decay for the low-energy part of the spectrum with a maximum, which shifts towards lower energies when increasing the time delay between two pulses. This clearly points towards a more complex physical picture than the self-similar model provides. Nevertheless, this simplified approach can still be successfully used to describe the high-energy part of the spectrum, i.e. beyond the observed maxima. These fits, showed as colored dashed lines in Fig. 3(b), reveal that the characteristic energy is the lowest at the time delay of 10 ns (red symbols) and has a value of $\varepsilon_0$=34.5 (0.9) eV, which is 30 times smaller compared to the single pulse measurement. Lower values of the characteristic energy for pulse pairs hint at lower electron temperature or ion charge state Z in comparison to the single pulse case. With a single pulse the laser light mainly interacts with the droplet, and due to its high density gets absorbed within a thin layer, leading to the generation of a hot plasma and consequently to the ejection of fast ions. For a pulse pair, the first weak pulse ablates material from a droplet, and the second pulse will therefore interact with this plasma as well as with the droplet. This second pulse may then get absorbed across a thicker layer, resulting in a colder plasma, in which the ion kinetic energies are reduced compared to the single-pulse case.

The presented experimental results on laser-produced tin plasmas demonstrate that by employing a picosecond pulse pair instead of a single pulse it is possible to greatly shift the ion kinetic energy distribution towards lower energies. Reduced kinetic energies make mitigation of the generated ions in LPPs by means of collisions with an ambient gas much more efficient [45]. By matching the laser beam size with the droplet size, further reduction in the number of fast ions should be achievable. Simultaneously, the recorded shadowgraphs showed that a picosecond pulse pair enables the tailoring of the target shape. Depending on the combination between the energy in the first pulse and a time delay between both pulses it is possible to obtain shapes similar to an acorn-like target, which in the interaction with a $CO_2$ main pulse may lead to a higher conversion efficiency into EUV light through the opening up of a larger parameter space for optimization.

**Figures:**

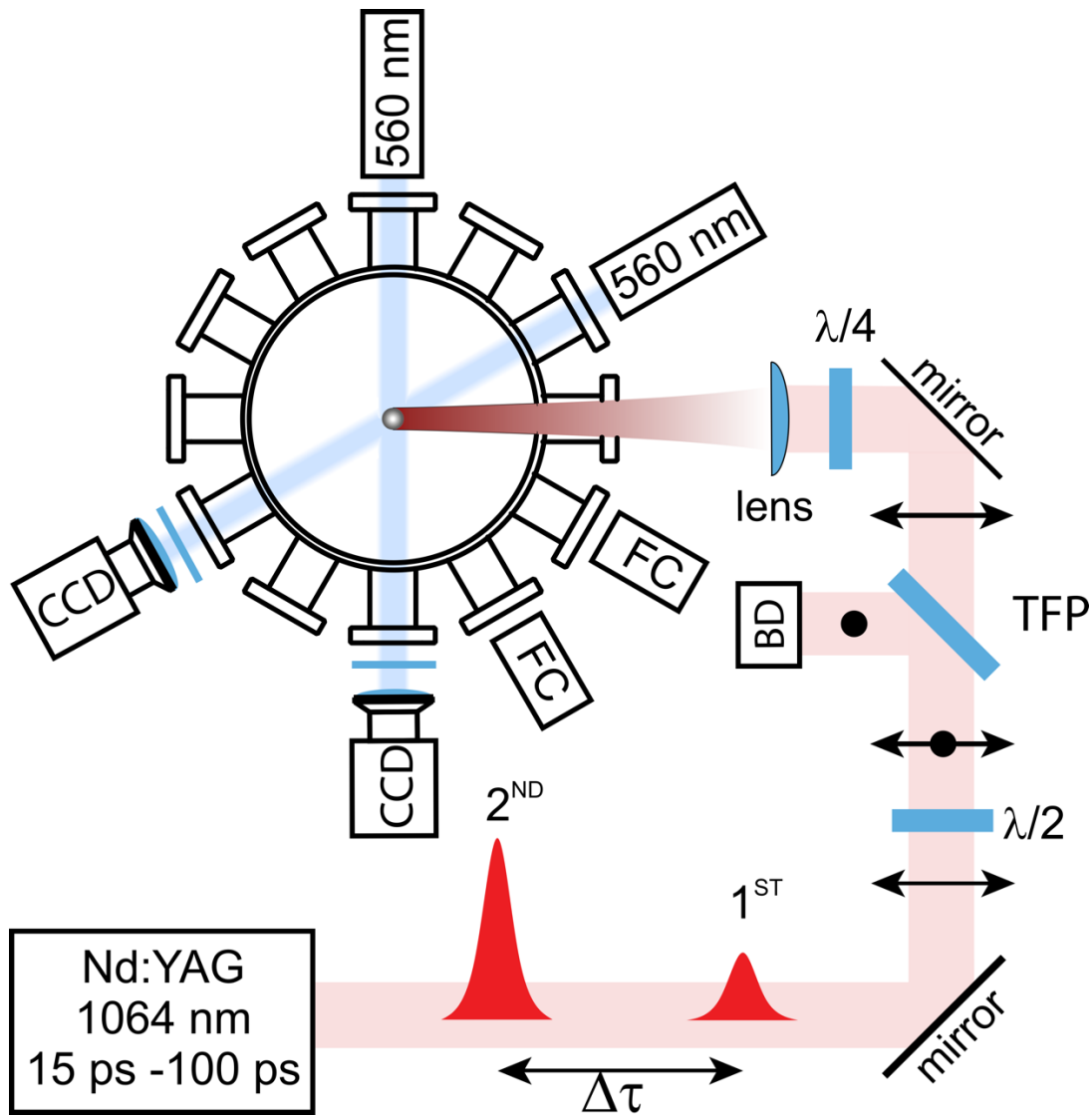

*Figure 1: A schematic representation of the experimental setup. A ps pulse pair is generated in a home-build Nd:YAG laser system with a controllable pulse duration (15 ps-100 ps) and a delay time Δτ between two pulses tunable from 0 to 1000 ns in increments of 10 ns. The pulse pair energy is set by means of a half-wave plate (λ/2) in combination with a thin film polarizer (TFP) and a beam dump (BD). Prior entering the vacuum chamber the polarization of the pulses is changed into circular with a quarter-wave plate (λ/4) and a convex lens (f = 600 mm) focuses pulses on a 30 µm tin droplet leading to the generation of plasma. The resulting ions are detected in time-of-flight measurements with Faraday cups (FC) positioned at 30° and 62° with respect to the laser plane. The time evolution of the droplet is captured on shadowgraphs obtained with two CCD cameras (at 90° and 150°) illumined by 560 nm light.*

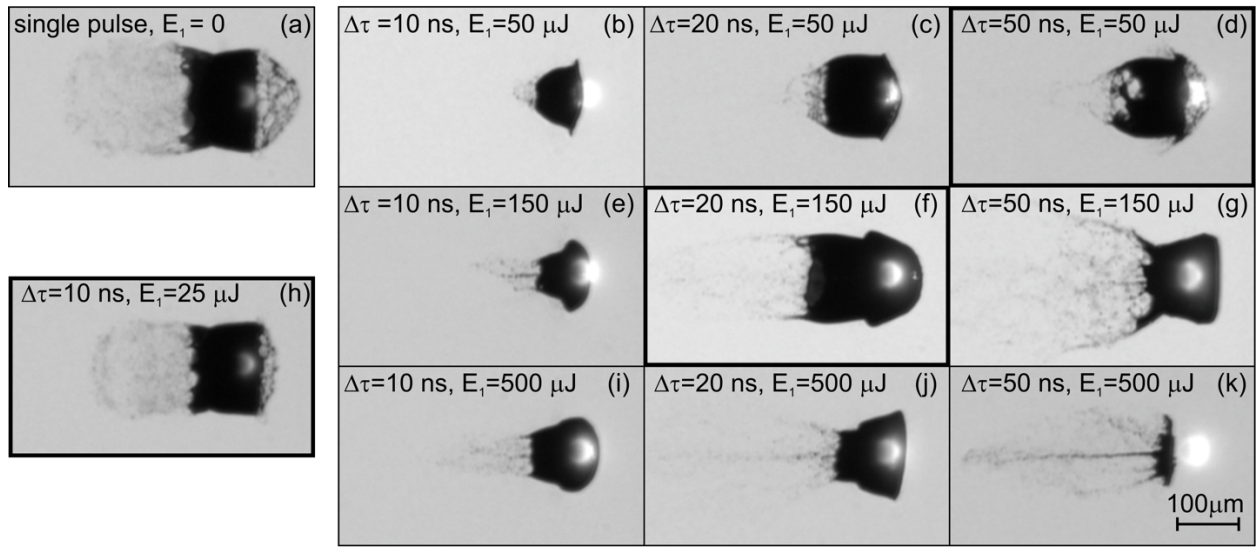

*Figure 2: Shadowgraphs showing the evolution of a 30 µm diameter tin droplet 550 ns after the interaction with a single pulse or a pulse pair for several typical parameter settings. The laser light hits the droplet from the right side and the bright spot is plasma light captured by a camera due to the long exposure time. (a) A 52 ps single pulse with energy of $E_2$ = 5 mJ. (b)-(d) A 52 ps pulse pair at various delay times Δτ with the 1st pulse energy set at $E_1$ = 50 µJ and the 2nd pulse energy fixed at $E_2$ = 5 mJ. (e)-(g) The same as (b)-(d), but with $E_1$ = 150 µJ. (h) A 52 ps pulse pair with a minimum energy in the 1st pulse of $E_1$ = 25 µJ, delayed by 10 ns with respect to the second pulse with energy of $E_2$ = 5 mJ. (i)-(k) The same as (b)-(d), but with $E_1$ = 500 µJ. Images highlighted with a thick frame ((d), (f) and (h)) resemble a typical acorn-like shape for which the highest CE has been reported [1].*

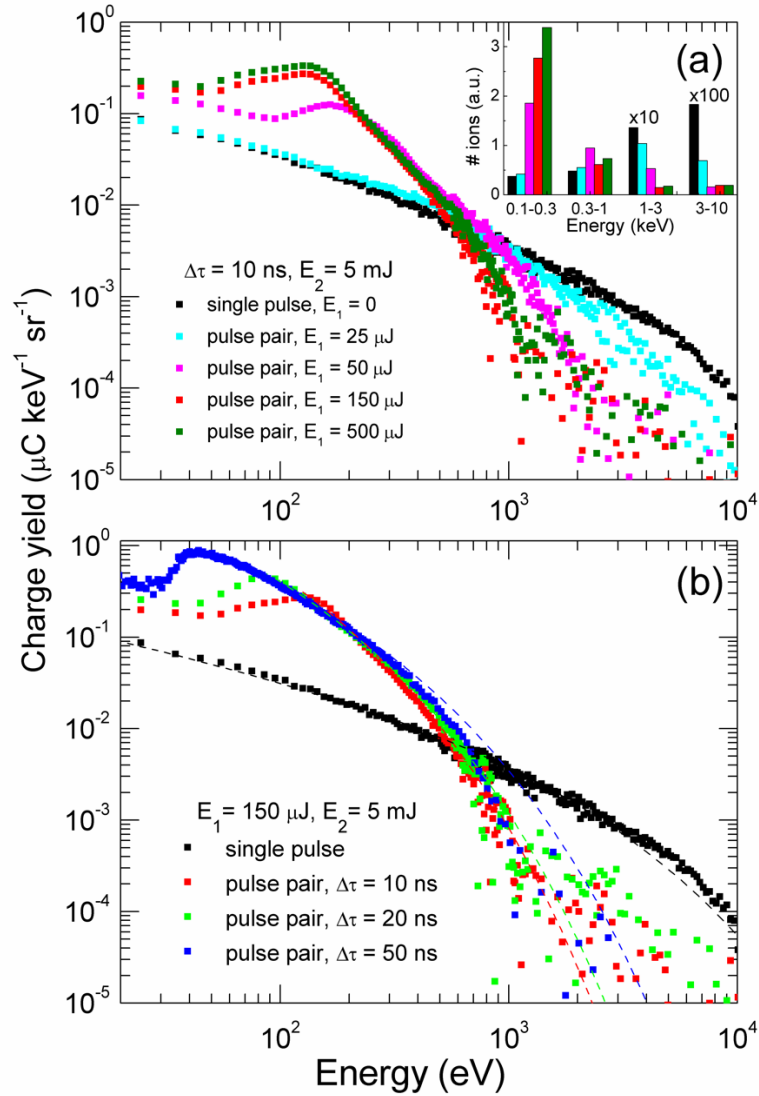

Figure 3: (a) Charge energy distributions measured by the 30° Faraday cup resulting from the ablation of a tin droplet by a single 5 mJ, 52 ps pulse (black symbols), and by 52 ps pulse pairs delayed by 10 ns with various energies in the first pulse ($E_1$) and the second pulse energy set at $E_2$ = 5 mJ (colored symbols). The inset shows the amount of collected ions obtained by integrating the energy distributions shown in (a) in four energy ranges. (b) The effect of different time delays $\Delta\tau$ between two pulses (colored symbols) on ions energy distributions with respect to a single pulse interaction (black symbols). The dashed lines are analytical fits to the distributions according to Eq. (2).